\documentstyle[12pt]{article}

\catcode`\@=11
\long\def\@makefntext#1{
\protect\noindent \hbox to 4.0pt {\hskip-.9pt
$^{{\ninerm\@thefnmark}}$\hfil}#1\hfill\relax}		

\def\@makefnmark{\hbox{$^{\@thefnmark}$}}  

\def\ps@myheadings{\let\@mkboth\@gobbletwo
\def\@oddhead{\hbox{}
\rightmark\hfil\ninerm\thepage}
\def\@oddfoot{}\def\@evenhead{\ninerm\thepage\hfil
\leftmark\hbox{}}\def\@evenfoot{}
\def\sectionmark##1{}\def\subsectionmark##1{}}

\expandafter\ifx\csname mathrm\endcsname\relax\def\mathrm#1{{\rm
#1}}\fi

\def\no{\nonumber\\}
\def\nl{\nonumber\\}
\def\nn{\nonumber}

\def\beq{\begin{equation}}
\def\eeq{\end{equation}}
\def\beqar{\begin{eqnarray}}
\def\eeqar{\end{eqnarray}}
\def\barr#1{\begin{array}{#1}}
\def\earr{\end{array}}

\def\bma{\begin{displaymath}}
\def\ema{\end{displaymath}}

\def\al{\alpha}
\def\be{\beta}
\def\Ga{\Gamma}
\def\ga{\gamma}
\def\de{\delta}
\def\De{\Delta}

\def\veps{\varepsilon}
\def\la{\lambda}
\def\om{\omega}
\def\si{\sigma}
\def\Si{\Sigma}

\def\refeq#1{\mbox{Eq. (\ref{#1})}}
\def\refeqs#1{\mbox{Eqs. (\ref{#1})}}
\def\refeqf#1{\mbox{(\ref{#1})}}
\def\refse#1{\mbox{section~\ref{#1}}}
\def\citere#1{\mbox{Ref.~\cite{#1}}}
\def\citeres#1{\mbox{Refs.~\cite{#1}}}
\def\@citere#1#2{\unskip\ Ref.~#1}
\def\citere#1{{\let\@cite\@citere\cite{#1}}}
\def\@citeres#1#2{\unskip\ Refs.~#1}
\def\citeres#1{{\let\@cite\@citeres\cite{#1}}}

\renewcommand{\L}{{\cal{L}}}

\def\mathswitchr#1{\relax\ifmmode{\mathrm{#1}}\else$\mathrm{#1}$\fi}
\newcommand{\Pf}{\mathswitch  f}
\newcommand{\PW}{\mathswitchr W}
\newcommand{\PZ}{\mathswitchr Z}

\newcommand{\PH}{\mathswitchr H}
\newcommand{\Pu}{\mathswitchr u}
\newcommand{\Pd}{\mathswitchr d}
\newcommand{\Pt}{\mathswitchr t}

\newcommand{\Ff}{\mathswitch f}
\newcommand{\Ffbar}{\mathswitch{\bar f}}
\newcommand{\FW}{\mathswitch W}
\newcommand{\FZ}{\mathswitch Z}
\newcommand{\FA}{\mathswitch A}
\newcommand{\FH}{\mathswitch H}

\def\mathswitch#1{\relax\ifmmode#1\else$#1$\fi}
\newcommand{\Mf}{\mathswitch {m_\Pf}}

\newcommand{\MW}{\mathswitch {M_\PW}}

\newcommand{\MZ}{\mathswitch {M_\PZ}}
\newcommand{\MH}{\mathswitch {M_\PH}}

\newcommand{\rw}{{\mathrm{W}}}
\newcommand{\sw}{\mathswitch {s_{\rw}}}
\newcommand{\cw}{\mathswitch {c_{\rw}}}

\newcommand{\Qf}{\mathswitch {Q_\Pf}}
\newcommand{\vf}{\mathswitch {v_\Pf}}
\newcommand{\af}{\mathswitch {a_\Pf}}

\newcommand{\se}{self-energy}
\newcommand{\ses}{self-energies}

\def\Vhat{\hat V}
\def\What{\hat W}
\def\Bhat{\hat B}
\def\Zhat{\hat Z}
\def\Ahat{\hat A}

\def\Hhat{\hat H}
\def\Phihat{\hat \Phi}
\def\phihat{\hat \phi}
\def\chihat{\hat \chi}

\def\xiQ{\xi_Q}
\def\xiB{\xi_B}
\def\rY{{\mathrm{Y}}}
\def\rT{{\mathrm{T}}}
\def\rS{{\mathrm{S}}}
\def\Uhat{\hat U}
\newcommand{\rL}{\mathswitchr L}
\newcommand{\rR}{\mathswitchr R}

\newcommand{\bfm}{BFM}
\renewcommand{\Re}{\mathop{\mathrm{Re}}}
\def\tr#1{\mathop{\mbox{tr}\left\{#1\right\}}}

\renewcommand{\thefootnote}{\fnsymbol{footnote}}

\newcounter{sectionc}\newcounter{subsectionc}\newcounter{subsubsectionc}
\renewcommand{\section}[1] {\vspace*{0.6cm}\refstepcounter{sectionc}%
\setcounter{subsectionc}{0}\setcounter{subsubsectionc}{0}\noindent
	{\normalsize\bf\thesectionc. #1}\par\vspace*{0.4cm}}
\renewcommand{\subsection}[1] {\vspace*{0.6cm}\addtocounter{subsectionc}{1}
	\setcounter{subsubsectionc}{0}\noindent
	{\normalsize\it\thesectionc.\thesubsectionc. #1}\par\vspace*{0.4cm}}
\renewcommand{\subsubsection}[1]
{\vspace*{0.6cm}\addtocounter{subsubsectionc}{1}
	\noindent {\normalsize\rm\thesectionc.\thesubsectionc.\thesubsubsectionc.
	#1}\par\vspace*{0.4cm}}

\newcounter{appendixc}
\newcounter{subappendixc}[appendixc]
\newcounter{subsubappendixc}[subappendixc]

\renewcommand{\appendix}[1] {\vspace*{0.6cm}
        \refstepcounter{appendixc}
        \setcounter{figure}{0}
        \setcounter{table}{0}
        \setcounter{equation}{0}
        \renewcommand{\thefigure}{\Alph{appendixc}.\arabic{figure}}
        \renewcommand{\thetable}{\Alph{appendixc}.\arabic{table}}
        \renewcommand{\theappendixc}{\Alph{appendixc}}
        \renewcommand{\theequation}{\Alph{appendixc}.\arabic{equation}}
        \noindent{\bf Appendix \theappendixc #1}\par\vspace*{0.4cm}}

\def\abstracts#1{{

\centering{\begin{minipage}{12.2truecm}\footnotesize\baselineskip=12pt\noindent
	\centerline{\footnotesize ABSTRACT}\vspace*{0.3cm}
	\parindent=0pt #1
	\end{minipage}}\par}}

\renewenvironment{thebibliography}[1]
	{\begin{list}{\arabic{enumi}.}
	{\usecounter{enumi}\setlength{\parsep}{0pt}
\setlength{\leftmargin 1.25cm}{\rightmargin 0pt}
	 \setlength{\itemsep}{0pt} \settowidth
	{\labelwidth}{#1.}\sloppy}}{\end{list}}

\newcounter{itemlistc}
\newcounter{romanlistc}
\newcounter{alphlistc}
\newcounter{arabiclistc}

\newcommand{\fcaption}[1]{
        \refstepcounter{figure}
        \setbox\@tempboxa = \hbox{\footnotesize Fig.~\thefigure. #1}
        \ifdim \wd\@tempboxa > 6in
           {\begin{center}
        \parbox{6in}{\footnotesize\baselineskip=12pt Fig.~\thefigure. #1}
            \end{center}}
        \else
             {\begin{center}
             {\footnotesize Fig.~\thefigure. #1}
              \end{center}}
        \fi}

\newcommand{\tcaption}[1]{
        \refstepcounter{table}
        \setbox\@tempboxa = \hbox{\footnotesize Table~\thetable. #1}
        \ifdim \wd\@tempboxa > 6in
           {\begin{center}
        \parbox{6in}{\footnotesize\baselineskip=12pt Table~\thetable. #1}
            \end{center}}
        \else
             {\begin{center}
             {\footnotesize Table~\thetable. #1}
              \end{center}}
        \fi}

\def\@citex[#1]#2{\if@filesw\immediate\write\@auxout
	{\string\citation{#2}}\fi
\def\@citea{}\@cite{\@for\@citeb:=#2\do
	{\@citea\def\@citea{,}\@ifundefined
	{b@\@citeb}{{\bf ?}\@warning
	{Citation `\@citeb' on page \thepage \space undefined}}
	{\csname b@\@citeb\endcsname}}}{#1}}

\newif\if@cghi
\def\cite{\@cghitrue\@ifnextchar [{\@tempswatrue
	\@citex}{\@tempswafalse\@citex[]}}
\def\citelow{\@cghifalse\@ifnextchar [{\@tempswatrue
	\@citex}{\@tempswafalse\@citex[]}}
\def\@cite#1#2{{$\null^{#1}$\if@tempswa\typeout
	{IJCGA warning: optional citation argument
	ignored: `#2'} \fi}}

 1
 1
 1

\font\ninerm=cmr9

\begin{document}

\thispagestyle{empty}
\def\thefootnote{\fnsymbol{footnote}}
\setcounter{footnote}{1}
\null
\renewcommand{\baselinestretch}{1}
\normalsize
\mbox{} \hfill BI-TP 95/17\\
\mbox{} \hfill WUE-ITP-95-010\\
\mbox{} \hfill hep-ph/9505271 \\
\mbox{} \hfill May 95
\vskip .5cm 
\vfill
\begin{center}
{\Large \bf
\boldmath{Gauge invariance, gauge-parameter independence \\
and properties of Green functions%
\footnote{To appear in the proceedings of the Ringberg Workshop
``Perspectives for electroweak interactions in
$\mathrm{e}^+\mathrm{e}^-$
collisions'',
Ringberg Castle, February 5-8, 1995, ed. B.A. Kniehl.}
}
\par} \vskip 2.5em
{\large
{\sc Ansgar Denner} \\[1ex]
{\normalsize \it Institut f\"ur Theoretische Physik, Universit\"at W\"urzburg\\
Am Hubland, D-97074 W\"urzburg, Germany}
\\[2ex]
{\sc Stefan Dittmaier, Georg Weiglein%
\footnote{E-mail: weiglein@physik.uni-bielefeld.de}%
\footnote{Supported by the Bundesministerium f\"ur Bildung und Forschung,
Bonn, Germany.} \\[1ex]
{\normalsize \it Theoretische Physik, Universit\"at Bielefeld\\
Postfach 10 01 31, D-33501 Bielefeld, Germany}
}
\par} \vskip 1em
\end{center} \par
\vskip .5cm 
\vfill
{\bf Abstract:} \par
The application of the background-field method to the
electroweak Standard Model is reviewed
and further explored. Special emphasis is
put on questions of gauge invariance and gauge-parameter
(in-)dependence. Owing to the gauge
invariance of the background-field effective action,
the vertex functions obey simple Ward identities
which imply important properties of the vertex
functions.
Carrying out the renormalization in a way respecting background-field
gauge invariance leads to considerable simplifications.
The generalization of the background-field method to the non-linear
realization of the scalar sector of the Standard Model is illustrated.
Furthermore, the interplay between gauge independence of the S-matrix
and Ward identities of vertex functions is investigated.
Finally, the Standard Model contributions to the $S$, $T$, and
$U$ parameters are
calculated and discussed within the background-field method.
\par
\null
\setcounter{page}{0}
\clearpage

\centerline{\normalsize\bf GAUGE INVARIANCE, GAUGE-PARAMETER
INDEPENDENCE}
\baselineskip=16pt
\centerline{\normalsize\bf AND PROPERTIES OF GREEN FUNCTIONS}

\vspace*{0.6cm}
\centerline{\footnotesize A.~DENNER}
\baselineskip=13pt
\centerline{\footnotesize\it Institut f\"ur Theoretische Physik, Universit\"at
W\"urzburg}
\baselineskip=12pt
\centerline{\footnotesize\it Am Hubland, D-97074 W\"urzburg, Germany}
\vspace*{0.3cm}
\centerline{\footnotesize and}
\vspace*{0.3cm}
\centerline{\footnotesize S.~DITTMAIER, G.~WEIGLEIN\footnote{Supported
by Bundesministerium f\"ur Bildung und Forschung, Bonn, Germany.}}
\baselineskip=13pt
\centerline{\footnotesize\it Theoretische Physik, Universit\"at
Bielefeld}
\baselineskip=12pt
\centerline{\footnotesize\it Postfach 10 01 31, D-33501 Bielefeld, Germany}

\vspace*{0.9cm}
\abstracts{The application of the background-field method to the
electroweak Standard Model is reviewed
and further explored. Special emphasis is
put on questions of gauge invariance and gauge-parameter
(in-)dependence. Owing to the gauge
invariance of the background-field effective action,
the vertex functions obey simple Ward identities
which imply important properties of the vertex
functions.
Carrying out the renormalization in a way respecting background-field
gauge invariance leads to considerable simplifications.
The generalization of the background-field method to the non-linear
realization of the scalar sector of the Standard Model is illustrated.
Furthermore, the interplay between gauge independence of the S-matrix
and Ward identities of vertex functions is investigated.
Finally, the Standard Model contributions to the $S$, $T$, and
$U$ parameters are
calculated and discussed within the background-field method.}

\normalsize\baselineskip=15pt
\setcounter{footnote}{0}
\renewcommand{\thefootnote}{\alph{footnote}}

\section{Introduction}
The properties
of gauge invariance and gauge-parameter independence, which
are inherent in all kinds of gauge theories, have recently gained
renewed interest.
The question of gauge dependence arises automatically
whenever physical observables, i.e.\ S-matrix elements, are not strictly
calculated order by order in perturbation theory. However, mixing
different orders of the perturbative expansion is sometimes unavoidable.
For example, the introduction of finite-width effects for unstable
particles or of running couplings can only be achieved by a resummation
of certain subsets of Feynman diagrams. Moreover, single off-shell
vertex functions have been parametrized by so-called form factors in the
literature. The physical significance of such objects is always
questionable. Every definition of quantities from
incomplete parts of S-matrix elements
(in a fixed order of perturbation theory) is necessarily
based on conventions but not on physical grounds.

At this point a few remarks on the difference between gauge invariance
and gauge-parameter independence are in order. Strictly speaking, one
can call only such objects gauge-invariant which are singlets with
respect to gauge transformations.
However, the gauge invariance of
the underlying Lagrangian has to be broken in order to quantize the
fields in perturbation theory. To this end a gauge-fixing term is added
to the Lagrangian depending on one or more free (gauge-)parameters,
which drop out in complete S-matrix elements.
If a quantity depends on the gauge parameters, it
depends on the gauge-fixing procedure.
On the other hand, it can be
shown that Green functions of gauge-invariant operators are independent
of the method of gauge fixing and thus gauge-parameter-independent.
However, the inverse conclusion is wrong in general: gauge-parameter
independence does {\it not} necessarily indicate gauge invariance.

The fact that the gauge-parameter dependence of individual vertex
functions (self-energies, vertex corrections, etc.) is
compensated within complete S-matrix elements motivated several authors to
rearrange the gauge-dependent parts between different vertex functions
resulting in definitions of
separately gauge-parameter-independent building blocks. Since such
splittings of vertex functions are not uniquely determined, different
proposals were made in the literature.  For example, in the context of
four-fermion processes different running couplings \cite{Ke89,Ku91} have been
defined. A more general procedure for eliminating the
gauge-parameter-dependent parts of vertex functions is given by the
so-called pinch technique (PT) \cite{PT1,Si92,PT2,PT3}. All these
approaches have
the common aim to define gauge-parameter-independent ``vertex
functions'' with improved theoretical properties.
In this context it should be noticed that these procedures are not free
of problems. The methods of \citeres{Ke89,Ku91} have no natural
generalization beyond four-fermion processes. On the other hand, the
application of the PT algorithm is not always clear, and the
universality (process independence) of the PT ``vertex functions''
has not rigorously been proven but only been
verified from examples \cite{PT2,PT3}.

Therefore, we pursue a completely different approach and study
directly consequences
of the underlying gauge invariance for vertex functions. The
background-field method (BFM) \cite{BFM,Ab81,Ab83} represents a well-suited
framework for such investigations. In the BFM the effective action,
which generates the vertex functions, is manifestly gauge-invariant, and
this invariance implies
simple Ward identities for vertex functions. For the calculation of
S-matrix elements, tree-like structures are formed with these vertex
functions, where the gauge fixing of the genuine tree part can be
chosen arbitrarily.

In \citeres{BFMvPT,bgflong} the BFM was applied to the electroweak Standard
Model (SM), and the consequences of the Ward identities were discussed. The
renormalization%
\footnote{The renormalization of the electroweak SM without fermions was
also discussed in \citere{Li}.} was carried out in a gauge-invariant way,
which led to considerable simplifications. Moreover,
it was shown that the Ward identities imply several improved properties
for vertex functions (compared to the conventional formalism)
concerning ultraviolet, infrared or high-energy behavior.
Furthermore, actual loop
calculations of S-matrix elements in general become
simpler using the BFM. This is mainly due to
the freedom of choosing the gauge for the tree parts independently from the
one in the loops.

The BFM brings together two important features: the gauge invariance of
the effective action and the clear distinction between classical and
quantum parts of fields. This fact renders the BFM well-suited
for integrating out heavy fields at one-loop level
directly in the path integral.
Firstly, the tree-level and one-loop effects can be isolated
in the path integral very easily. Secondly, choosing a
definite gauge (e.g.\ the unitary gauge) for the background fields
drastically simplifies intermediate steps in the $1/M$-expansion for
the heavy field of mass $M$.
Thus, a one-loop effective Lagrangian can be directly derived and
by inverting the transformation to the
definite background gauge after the $1/M$-expansion one recovers a
manifestly gauge-invariant result.
Such a procedure was worked out in \citere{HHint}
and applied to an $\mathrm{SU}(2)_{\rw}$ gauge theory and the SM.

In \citere{BFMvPT} it was realized that the building blocks
obtained within the PT in QCD%
\footnote{In QCD this fact was also pointed out in \citere{Ha94}.}
and the SM coincide
with the corresponding BFM vertex functions in 't~Hooft--Feynman gauge.
This observation and further investigations in \citere{BFMvPT} clarified
the origin of certain desirable properties \cite{Si92} noticed for the PT
``vertex functions''.
In particular, in the BFM the QED-like Ward identities are derived from
gauge invariance and imply all the other properties as shown in
\citere{BFMvPT}.  Since this is true in the BFM for arbitrary
gauges of the quantum fields, the above-mentioned
desirable properties are related to the gauge invariance of the effective
action rather than to
the absence of gauge-parameter dependence.

In this article we first review the basic results of the
application of the BFM to the SM.
They have been worked out in \citere{bgflong}
for the usual linear realization of the scalar Higgs doublet.
The generalization of the BFM to the non-linear realization of the
scalar sector was described in \citere{HHint}.
We further explore the connection between
gauge-parameter-independent formulations and the BFM.
Finally, we focus on the SM contributions to the
$S$, $T$, and $U$ parameters, which have been originally introduced
\cite{STU} in order to
quantify new-physics effects beyond the SM entering via vacuum
polarization only.
Comparing the BFM results for the $S$, $T$, and $U$ parameters
with the ones obtained within the PT \cite{DKS},
the relevance of the latter is discussed.

{\samepage
The article is organized as follows: in \refse{se:bfm} we review the
application of the background field method to the electroweak SM.
The renormalization of the SM in the BFM is discussed in \refse{se:ren}.
In \refse{se:nonlin} we summarize the virtues of the BFM in the formulation
of the SM with a non-linear realization of the Higgs sector. In
\refse{se:GIVvGIP} we elaborate on
the connection between gauge-parameter independence of vertex functions
and symmetry relations.
As a further illustration we treat the $S$, $T$, and $U$ parameters in
the BFM in \refse{se:STU}.
}

\pagebreak
\section{The Background-Field Method for the Electroweak Standard Model}
\label{se:bfm}
\vspace{-1em}
\subsection{The Construction of the gauge-invariant Effective Action}
The background-field method~\cite{BFM,Ab81} (\bfm) is a technique for
quantizing gauge theories without losing explicit gauge invariance
of the effective action.
This is done by decomposing the usual fields $\hat\varphi$ in the classical
Lagrangian $\L_{\mathrm{C}}$ into background fields $\hat\varphi$
and quantum fields $\varphi$,
\beq
\label{eq:quantLc}
{\cal L}_{\mathrm{C}}(\hat\varphi) \rightarrow
{\cal L}_{\mathrm{C}}(\hat\varphi + \varphi).
\eeq
While the background fields are treated as external sources, only
the quantum fields are
variables of integration in the functional integral.
A gauge-fixing term is added which breaks only the
invariance with respect to quantum gauge transformations
but retains the  invariance of the functional integral with respect to
background-field gauge transformations.
{}From the functional integral an effective action $\Ga[\hat\varphi]$ for
the background fields is
derived which is invariant under gauge transformations of the
background fields and thus gauge-invariant.

The S-matrix is constructed by forming trees with vertex functions
from $\Gamma[\hat\varphi]$
joined by background-field propagators. These
propagators are defined by adding a
gauge-fixing term to $\Gamma[\hat\varphi]$.
This gauge-fixing term
is only relevant for the construction of connected Green
functions and S-matrix elements.
It is not
related to the term used to fix the gauge inside loop diagrams, i.e.\ in
the functional integral, and the associated gauge parameters $\xiB^i$
only enter tree level quantities but not the higher-order contributions
to the vertex functions.
In particular, in linear
background gauges only the tree-level propagators are
affected by the background gauge fixing.
The equivalence of the S-matrix in the \bfm\
to the conventional one has been proven in \citeres{Ab83,Re85,Be94}.

For our discussion of the SM we use the conventions of
\citeres{bgflong,Dehab}.
The complex scalar SU$(2)_{\rw}$ doublet field of the minimal
Higgs sector is written as the sum of a
background Higgs field $\hat\Phi$ having the usual non-vanishing
vacuum expectation value $v$, and a quantum Higgs field
$\Phi$ whose vacuum expectation value is zero:
\begin{equation}
\hat\Phi(x) = \left( \begin{array}{c}
\phihat ^{+}(x) \\ \frac{1}{\sqrt{2}}\bigl(v + {\hat H}(x) +i
\chihat (x) \bigr)
\end{array} \right) , \qquad
\Phi (x) = \left( \begin{array}{c}
\phi ^{+}(x) \\ \frac{1}{\sqrt{2}}\bigl(H(x) +i \chi (x) \bigr)
\end{array} \right) .
\label{eq:Hbq}
\end{equation}
Here ${\hat H}$ and $H$ denote the physical background and
quantum Higgs field, respectively, and $\phihat ^{+}, \chihat,
\phi ^{+}, \chi$
are the unphysical Goldstone fields.

The generalization of the 't~Hooft gauge fixing to the BFM~\cite{Sh81} reads
\beqar\label{tHgf}
\L_{\mathrm{GF}} &=& {}- \frac{1}{2\xiQ^W}
\biggl[(\de^{ac}\partial_\mu + g_2 \veps^{abc}\hat
W^b_\mu)W^{c,\mu}
       -ig_2\xiQ^W\frac{1}{2}(\hat\Phi^\dagger_i
\si^a_{ij}\Phi_j
                  - \Phi^\dagger_i
\si^a_{ij}\hat\Phi_j)\biggr]^2 \no
                 && {}- \frac{1}{2\xiQ^B}
\biggl[\partial_\mu B^{\mu}
       +ig_1\xiQ^B\frac{1}{2}(\hat\Phi^\dagger_i \Phi_i
                  - \Phi^\dagger_i \hat\Phi_i)\biggr]^2,
\eeqar
where  $\What_{\mu }^{a}$, $a$=1,2,3, represents the triplet of gauge
fields associated with the weak isospin group SU$(2)_{\rw}$,
and $\Bhat_{\mu }$ the gauge field  associated with the group
U$(1)_{\rY}$ of weak hypercharge $Y_{\rw}$.
The Pauli matrices are denoted by $\si^a$, $a=1,2,3$,
and $\xiQ^W$,
$\xiQ^B$ are parameters associated with the gauge fixing of the
quantum fields, one for SU$(2)_{\rw}$ and one for U$(1)_{\rY}$.
In order to avoid tree-level mixing between the quantum $\FA$ and $\FZ$
fields, we set $\xiQ=\xiQ^W=\xiQ^B$ in the following.
Background-field gauge invariance implies that the background gauge
fields appear only within a covariant derivative in the gauge-fixing term and
that the terms in brackets transform according to the adjoint
representation of the gauge group. The
gauge-fixing term (\refeq{tHgf}) translates to the conventional one upon
replacing the background Higgs field by its vacuum expectation value and
omitting the background SU$(2)_{\rw}$ triplet field $\hat W^a_\mu$.

The vertex functions can be calculated directly from Feynman rules that
distinguish between quantum and
background fields. Whereas the quantum fields appear only inside
loops, the background fields are associated with external lines.
Apart from doubling of the gauge and Higgs fields, the
BFM Feynman rules differ from the conventional ones
only owing to the gauge-fixing and ghost terms.
Because these terms are quadratic in the quantum fields, they affect
only vertices that involve exactly two quantum fields and additional
background fields.
Since the gauge-fixing term is non-linear in the fields, the
gauge parameter enters also the gauge-boson vertices.
The fermion fields are treated as usual,
they have the conventional Feynman rules,
and no distinction needs to be made between external and internal fields.
A complete set of BFM Feynman rules for the electroweak SM has
been given in~\citere{bgflong}.

Despite the distinction between background
and quantum fields, calculations in the BFM become in general simpler
than in the conventional formalism.
This is in
particular the case in the 't~Hooft--Feynman gauge ($\xiQ=1$)
for the quantum fields where many vertices simplify a lot.
Moreover, the gauge fixing of the background fields is totally unrelated
to the gauge fixing of the quantum fields\cite{Re85}.
This freedom can be
used to choose a particularly suitable background gauge,
e.g.~the unitary gauge. 
In this way the number of Feynman diagrams can  considerably be reduced.

\subsection{Ward Identities} 
\label{sec:WI}

As can be directly read off from Eqs.~(21), (22) of \citere{bgflong},
the invariance of the effective action under the
background gauge transformations yields
\newcommand{\dgd}[1]{\frac{\de\Ga}{\de#1}}
\beqar
0 = \frac{\de\Ga}{\de\hat\theta^\FA} &=&
-\partial_\mu\dgd{\hat\FA_\mu}
- ie \biggl( \hat\FW^+_\mu\dgd{\hat\FW^+_\mu}
- \hat\FW^-_\mu\dgd{\hat\FW^-_\mu} \biggr)
- ie \biggl( \hat\phi^+\dgd{\hat\phi^+}
- \hat\phi^-\dgd{\hat\phi^-} \biggr)   \nl
&& {}
+ ie \sum_f Q_f\biggl(\bar f \dgd{\bar f} + \dgd{f} f \biggr),\\
0 = \frac{\de\Ga}{\de\hat\theta^\FZ} &=&
-\partial_\mu\dgd{\hat\FZ_\mu}
+ ie\frac{\cw}{\sw}
\biggl(\hat\FW^+_\mu\dgd{\hat\FW^+_\mu}
-\hat\FW^-_\mu\dgd{\hat\FW^-_\mu}\biggr)
\nl && {}
   + ie\frac{\cw^2-\sw^2}{2\cw\sw}
       \biggl(\hat\phi^+\dgd{\hat\phi^+}
     - \hat\phi^-\dgd{\hat\phi^-} \biggr)
   - e\frac{1}{2\cw\sw}
       \biggl((v+\hat\FH)\dgd{\hat\chi}
     - \hat\chi\dgd{\hat\FH} \biggr) \nl
&& {}      - ie \sum_f \biggl(\bar f (v_f+a_f\ga_5) \dgd{\bar f}
                          + \dgd{f} (v_f-a_f\ga_5) f \biggr)
,\\
0 = \frac{\de\Ga}{\de\hat\theta^\pm} &=&
-\partial_\mu\dgd{\hat\FW^\pm_\mu}
\mp ie \hat\FW^\mp_\mu \biggl(\dgd{\hat\FA_\mu}
      -\frac{\cw}{\sw}\dgd{\hat\FZ_\mu}\biggr)
\pm ie (\hat\FA_\mu - \frac{\cw}{\sw}\hat\FZ_\mu)\dgd{\hat\FW^\pm_\mu}
\nl && {}
\mp ie \frac{1}{2\sw} \hat\phi^\mp \biggl(\dgd{\hat\FH}
             \pm i\dgd{\hat\chi}\biggr)
\pm ie \frac{1}{2\sw} (v + \hat\FH \pm i \hat\chi) \dgd{\hat\phi^\pm} \nl
&&{}         - ie\frac{1}{\sqrt{2}\sw} \sum_{(f_+,f_-)} \biggl(
\bar f_\pm \frac{1 +\ga_5}{2} \dgd{\bar f_\mp}
                          + \dgd{f_\pm} \frac{1-\ga_5}{2} f_\mp \biggr),
\label{eq:inv}
\eeqar
where $\vf = (I^3_{\rw,f} - 2\sw^2 Q_f)/(2\sw\cw)$
and $\af = I^3_{\rw,f}/(2\sw\cw)$.
In \refeq{eq:inv}
$f_\pm$ denote the fermions with isospin $\pm1/2$, and the sum
in the last line runs over all isospin doublets.
The electric unit charge is denoted by $e$ as usual, and the Weinberg
angle $\theta_\PW$ is fixed by the mass ratio,
\beq
\cw =\cos\theta_{\rw}= \frac{\MW}{\MZ},
\qquad \sw =\sin\theta_{\rw} = \sqrt{1-\cw^2}.
\eeq
By differentiating \refeq{eq:inv} with respect to background fields
and setting the fields equal to zero,
one obtains Ward identities for the vertex function that
are precisely the Ward identities related to the classical Lagrangian.
This is in contrast to the conventional formalism where, owing to
the gauge-fixing procedure, explicit gauge invariance is lost,
and Ward identities are obtained from
invariance under BRS transformations.
These Slavnov--Taylor identities have a more complicated
structure and in general involve ghost contributions.

The BFM Ward identities are valid in all orders of perturbation theory
and hold for arbitrary values of the quantum gauge parameter $\xi_Q$.
They relate one-particle irreducible Green functions.
In particular, the two-point functions do not contain tadpole
contributions. These appear explicitly in the Ward identities.

For illustration and later use, we list some of the Ward identities.
Concerning the notation and conventions for the vertex
functions we follow \citere{bgflong} throughout.
The two-point
functions fulfill the following Ward identities:
\beqar
\label{eq:sega}
k^\mu \Ga^{\Ahat\Ahat}_{\mu\nu}(k) = 0,
\parbox{3cm}{\hfill $k^\mu \Ga^{\Ahat\Zhat}_{\mu\nu}(k)$}
&=& 0, \\
\label{eq:segachi}
k^\mu \Ga^{\Ahat\Hhat}_{\mu}(k)  = 0,
\parbox{3cm}{\hfill $k^\mu \Ga^{\Ahat\chihat}_{\mu}(k) $}
&=& 0, \\
\label{eq:seZ1}
k^\mu \Ga^{\Zhat\Zhat}_{\mu\nu}(k) -i\MZ \Ga^{\chihat\Zhat}_\nu(k)
&=& 0, \\
\label{eq:seZ2}
k^\mu \Ga^{\Zhat\chihat}_{\mu}(k) -i\MZ \Ga^{\chihat\chihat}(k)
+\frac{ie}{2\sw\cw} \Ga^{\Hhat}(0) &=& 0 , \\
\label{eq:seW1}
k^\mu \Ga^{\What^\pm\What^\mp}_{\mu\nu}(k)
\mp\MW \Ga^{\phihat^\pm\What^\mp}_\nu(k) &=& 0, \\
\label{eq:seW2}
k^\mu \Ga^{\What^\pm\phihat^\mp}_{\mu}(k)
\mp\MW \Ga^{\phihat^\pm\phihat^\mp}(k)
\pm\frac{e}{2\sw} \Ga^{\Hhat}(0)
&=& 0.
\eeqar

The Ward identities for the gauge-boson--fermion vertices read
\beqar
\label{WIAff}
\!\!&& k^\mu \Ga^{\Ahat\Ffbar\Ff}_{\mu}(k,\bar p, p) = -e\Qf
[\Ga^{\Ffbar\Ff}(\bar p) - \Ga^{\Ffbar\Ff}(-p)],  \\
\label{WIZff}
\!\!&& k^\mu \Ga^{\Zhat\Ffbar\Ff}_{\mu}(k,\bar p,p) - i
\MZ \Ga^{\hat\chi\Ffbar\Ff}(k,\bar p,p) = e
[\Ga^{\Ffbar\Ff}(\bar p)(\vf-\af\ga_5) 
- (\vf+\af\ga_5) \Ga^{\Ffbar\Ff}(-p)], \\
\label{WIWff}
\!\!&& k^\mu \Ga^{\What^\pm\Ffbar_\pm\Ff_\mp}_{\mu}(k,\bar p,p) \mp
\MW \Ga^{\phihat^\pm\Ffbar_\pm\Ff_\mp}(k,\bar p,p) =
\frac{e}{\sqrt{2}\sw}
[\Ga^{\Ffbar_\pm\Ff_\pm}(\bar p)\om_- - \om_+ \Ga^{\Ffbar_\mp\Ff_\mp}(-p)] .
\hspace{2em}
\eeqar
The triple-gauge-boson vertices obey
\beqar
\label{WIAWW}
k^\mu \Ga^{\Ahat\What^+\What^-}_{\mu\rho\si}(k,k_+,k_-) &=&
e [\Ga^{\What^+\What^-}_{\rho\si}(k_+) -
\Ga^{\What^+\What^-}_{\rho\si }(-k_-)], \\
k_+^\rho \Ga^{\hat\FA\hat\FW^+\hat\FW^-}_{\mu\rho\si}(k,k_+,k_-)
&-&
  \MW \Ga^{\hat\FA\hat\phi^+\hat\FW^-}_{\mu\si}(k,k_+,k_-) = \no
&& \quad +e\left[\Ga^{\hat\FW^+\hat\FW^-}_{\mu\si}(-k_-) -
  \Ga^{\hat\FA\hat\FA}_{\mu\si}(k)
  +\frac{\cw}{\sw} \Ga^{\hat\FA\hat\FZ}_{\mu\si}(k)\right] ,
\label{WIWAW}\\
k_-^\si \Ga^{\hat\FA\hat\FW^+\hat\FW^-}_{\mu\rho\si}(k,k_+,k_-)
&+&
  \MW \Ga^{\hat\FA\hat\FW^+\hat\phi^-}_{\mu\rho}(k,k_+,k_-) = \no
&& \quad -e\left[\Ga^{\hat\FW^-\hat\FW^+}_{\mu\rho}(-k_+) -
  \Ga^{\hat\FA\hat\FA}_{\mu\rho}(k)
  +\frac{\cw}{\sw} \Ga^{\hat\FA\hat\FZ}_{\mu\rho}(k)\right] ,
\label{WIWWA}
\\
\label{WIZWW}
k^\mu \Ga^{\Zhat\What^+\What^-}_{\mu\rho\si}(k,k_+,k_-) &-&
i\MZ \Ga^{\chihat\What^+\What^-}_{\rho\si}(k,k_+,k_-)  = \nl
&& \quad -e\frac{\cw}{\sw} [\Ga^{\What^+\What^-}_{\rho\si}(k_+) -
\Ga^{\What^+\What^-}_{\rho\si }(-k_-)], \\
k_+^\rho \Ga^{\hat\FZ\hat\FW^+\hat\FW^-}_{\mu\rho\si}(k,k_+,k_-)
&-&
  \MW \Ga^{\hat\FZ\hat\phi^+\hat\FW^-}_{\mu\si}(k,k_+,k_-) = \no
&& \quad -e\frac{\cw}{\sw}\left[\Ga^{\hat\FW^+\hat\FW^-}_{\mu\si}(-k_-) -
  \Ga^{\hat\FZ\hat\FZ}_{\mu\si}(k)
  +\frac{\sw}{\cw} \Ga^{\hat\FZ\hat\FA}_{\mu\si}(k)\right] ,
\label{WIWZW}\\
k_-^\si \Ga^{\hat\FZ\hat\FW^+\hat\FW^-}_{\mu\rho\si}(k,k_+,k_-)
&+&
  \MW \Ga^{\hat\FZ\hat\FW^+\hat\phi^-}_{\mu\rho}(k,k_+,k_-) = \no
&& \quad +e\frac{\cw}{\sw}\left[\Ga^{\hat\FW^-\hat\FW^+}_{\mu\rho}(-k_+) -
  \Ga^{\hat\FZ\hat\FZ}_{\mu\rho}(k)
  +\frac{\sw}{\cw} \Ga^{\hat\FZ\hat\FA}_{\mu\rho}(k)\right] .
\hspace{2em}
\label{WIWWZ}
\eeqar
Note that the Ward identities involving only fermions and photons are
exactly those of QED.

\section{Renormalization of the Standard Model}
\label{se:ren}
\vspace{-1em}
\subsection{Impact of Gauge Invariance on Renormalization}
The BFM gauge invariance has
important consequences for the structure of the
renormalization constants necessary to render Green functions
and S-matrix elements finite.
The arguments which we give in the following are made
explicit for the one-loop level.%
\footnote{We implicitly assume the existence of an invariant
regularization scheme.}
It is easy, however, to extend them by induction to arbitrary
orders in perturbation theory. Because the renormalization of the
fermionic sector is similar to the one
in the conventional formalism, we leave
it out%
\footnote{It is included in \citere{bgflong}.}.

We introduce the following renormalization constants for the parameters:
\beqar
e_0 &=& Z_e e = (1 + \delta Z_e) e , \no
{\MW^2}_{,0} &=& \MW^2 + \delta \MW^2 , \qquad
{\MZ^2}_{,0} = \MZ^2 + \delta \MZ^2 , \qquad
{\MH^2}_{,0} = \MH^2 + \delta \MH^2 , \no
t_0 &=& t + \delta t .
\label{eq:renconsts1}
\eeqar
The tadpole counterterm $\delta t$ renormalizes the term
in the Lagrangian linear
in the Higgs field $\Hhat$, which we denote by $t \Hhat(x)$ with
$t = v (\mu^2 - \la v^2/4)$.
It corrects for the shift in the minimum of the Higgs potential
due to radiative corrections. Choosing $v$ as the correct vacuum
expectation value of the Higgs field $\Phihat$ is equivalent to
the vanishing of $t$.
In principle, the renormalization constant $\de t$ is not necessary,
and one could work
with arbitrary or even without tadpole renormalization. In these cases,
however, one would have to take into account explicit tadpole contributions.

Following the QCD treatment of~\citere{Ab81},
we introduce field renormalization only for the background fields
\beqar
\What_{0}^{\pm}  & = & Z_{\What}^{1/2} \What^{\pm}
  = (1+\frac{1}{2}\delta Z_{\What}) \What^{\pm} , \no
\left(\barr{l} \Zhat_{0} \\ \Ahat_{0} \earr \right)  & = &
\left(\barr{ll} Z_{\Zhat\Zhat}^{1/2} & Z_{\Zhat\Ahat}^{1/2}  \\[1ex]
                Z_{\Ahat\Zhat}^{1/2} & Z_{\Ahat\Ahat}^{1/2}
      \earr
\right)
\left(\barr{l} \Zhat \\ \Ahat \earr \right)   =
\left(\barr{cc} 1 + \frac{1}{2}\delta Z_{\Zhat\Zhat} &
\frac{1}{2}\delta Z_{\Zhat\Ahat} \\ [1ex]
\frac{1}{2}\delta Z_{\Ahat\Zhat}  & 1 + \frac{1}{2}\delta
Z_{\Ahat\Ahat}
\earr \right)
\left(\barr{l} \Zhat \\[1ex] \Ahat \earr \right)  , \no
\Hhat_{0} & = & Z_{\Hhat}^{1/2} \Hhat
 = (1+\frac{1}{2}\delta Z_{\Hhat}) \Hhat, \no
\chihat_{0} & = & Z_{\chihat}^{1/2} \chihat =
  (1+\frac{1}{2}\delta Z_{\chihat}) \chihat , \no
\phihat_{0}^{\pm} & = & Z_{\phihat}^{1/2} \phihat^{\pm} =
  (1+\frac{1}{2}\delta Z_{\phihat}) \phihat^{\pm} .
\label{eq:renconsts2}
\eeqar

In order to preserve the background-field gauge invariance,
the renormalized effective action has to be invariant under
background-field gauge transformations.
This restricts the possible counterterms and relates the
renormalization constants introduced above. These relations can be
derived from the requirement that the
renormalized vertex functions fulfill Ward identities of the same form
as the unrenormalized ones.
As a consequence, also the counterterms have to fulfill these Ward
identities. An analysis of the Ward identities yields~\cite{bgflong}:
\beqar
\label{eq:delZB}
\delta Z_{\Ahat\Ahat} &=& - 2 \delta Z_e, \qquad
\delta Z_{\Zhat\Ahat} = 0, \qquad
\delta Z_{\Ahat\Zhat} = 2 \frac{\cw}{\sw}
    \frac{\delta \cw ^2}{\cw ^2} , \no
\delta Z_{\Zhat\Zhat} &=& - 2 \delta Z_e -
    \frac{\cw ^2 - \sw ^2}{\sw^2} \frac{\delta \cw ^2}{\cw ^2} , \qquad
\delta Z_{\What} = - 2 \delta Z_e -
    \frac{\cw ^2}{\sw^2} \frac{\delta \cw ^2}{\cw ^2} , \no
\delta Z_{\Hhat} &=& \delta Z_{\chihat} = \delta Z_{\phihat} 
      = - 2 \delta Z_e -
	\frac{\cw ^2}{\sw^2} \frac{\delta \cw ^2}{\cw ^2} +
	\frac{\delta \MW^2}{\MW^2} ,
\eeqar
where
\beq
\frac{\delta \cw ^2}{\cw ^2} =
\frac{\delta \MW^2}{\MW^2} - \frac{\delta \MZ^2}{\MZ^2} .
\eeq

The relations \refeq{eq:delZB} express the
field renormalization constants of all gauge bosons and scalars
completely in terms of the renormalization constants of the
electric charge and the particle masses.
With this set of renormalization constants all background-field vertex
functions become finite%
\footnote{Beyond one-loop
order one needs in addition a renormalization of
the quantum gauge parameters~\cite{Ab81}. At one-loop
level these counterterms
do not enter the background-field vertex functions because
$\xi_Q$ does not appear in pure background-field vertices.
Clearly, the renormalization of gauge parameters is irrelevant
for
gauge-independent quantities such as
S-matrix elements at any order.}.
This is evident since the divergences of the vertex functions are
subject to the same restriction as the counterterms.
In \citere{bgflong} it has been verified explicitly at one-loop
order that
a renormalization based on the on-shell definition of all
parameters can consistently be used in the BFM. It
renders all vertex functions finite while respecting the full
gauge symmetry of the BFM.

As the field renormalization constants are fixed by \refeq{eq:delZB},
the propagators in general acquire residues
being different from unity but finite.
This is similar to the minimal on-shell
scheme of the conventional formalism\cite{BHS}
and has to be corrected in the S-matrix
elements by UV-finite wave-function renormalization constants.
However, just as in QED, the on-shell definition of the electric
charge together with gauge invariance automatically fixes the residue
of the photon propagator to unity.

As a consequence of the relations between the renormalization constants,
the counterterm vertices of the background fields
have a much simpler structure than the ones in the conventional
formalism (see e.g.~\citere{Dehab}).
In fact, all vertices originating from a separately gauge-invariant term
in the Lagrangian acquire the same renormalization constants.
The explicit form of the counterterm vertices
at one-loop order has been given in~\citere{bgflong}.

\subsection{Gauge-Parameter Independence of Counterterms and Running
Couplings}
If the renormalized
parameters are identified with the physical electron charge and
the physical particle masses, they are manifestly
gauge-independent. Moreover, the original bare parameters in the
Lagrangian are obviously
gauge-independent, as they represent free parameters of the
theory. The same is true for the bare charge
and the bare weak mixing angle as these are directly related to the free
bare parameters. Consequently, the counterterms
$\delta Z_e$ and $\delta\cw^2$ for the gauge couplings are
gauge-independent. The relations
\refeq{eq:delZB} therefore imply that the field renormalizations
of all gauge-boson fields are gauge-independent. This is in
contrast to the conventional formalism where the field
renormalizations in the on-shell scheme are gauge-dependent.

In contrast to
$\delta Z_e$ and $\delta\cw^2$ the mass counterterms
are not gauge-independent. The bare masses depend on the bare
vacuum expectation value $v_0$ of the Higgs field,
which is not a free parameter of the theory.
Whereas the renormalized value $v=2\sw\MW/e$ is
gauge-independent, the bare quantity $v_0$ and the corresponding
counterterm $\delta v$ are not. 
As a consequence, the bare masses are gauge-dependent.
Thus, the counterterms
$\delta\MW^2$, $\delta\MZ^2$, $\delta\MH^2$, $\delta\Mf$ and
$\delta t$ are also gauge-dependent. The physical
masses, however, are determined by the pole positions of the
propagators, i.e.~the zeros of
$k^2 - M^2 - \delta M^2 + C \delta t/\MH^2 + \Sigma(k^2) +
C T^{\Hhat}/\MH^2$,
where $C$ denotes the coupling of the fields to the Higgs field
and $\Sigma(k^2)$ the relevant \se.
The linear combination $\delta M^2 - C \delta t/\MH^2$ of the
mass and tadpole counterterm, however, is
independent of $\delta v$ and thus gauge-independent%
\footnote{Note that the mass counterterms become gauge-independent
if one chooses $\de t=0$.}.

Just as in QED, one can define running couplings in the BFM for the SM
via na{\"\i}ve Dyson summation of \ses\ as follows:
\beqar
\label{eq:runcoupl}
e^2(q^2) &=& \frac{e_0^2}{1 + \Re \Pi^{\Ahat\Ahat}_0(q^2)}
         = \frac{e^2}{1 + \Re \Pi^{\Ahat\Ahat}(q^2)} , \no
g_2^2(q^2) &=& \frac{g_{2,0}^2}{1 + \Re\Pi^{\What\What}_0(q^2)}
         = \frac{g_2^2}{1 + \Re\Pi^{\What\What}(q^2)} ,
\eeqar
where
\beq
g_{2,0} = \frac{e_0}{s_{\rw,0}} \quad\mbox{and}\quad g_2 = \frac{e}{\sw},
\eeq
and the subscript
``0'' denotes bare quantities.
The quantities $\Pi^{\Vhat\Vhat'}$ are related to the transverse parts of
the gauge-boson \ses\ as follows:
\beq
\Pi^{\Vhat\Vhat'}(q^2) =
\frac{\Sigma^{\Vhat\Vhat'}_\rT(q^2)-\Sigma^{\Vhat\Vhat'}_\rT(0)}{q^2}.
\eeq
The relations \refeq{eq:delZB} give rise to a number of
nice properties
of the running couplings. As indicated
in \refeq{eq:runcoupl}, the renormalization constants
cancel. Consequently, the running couplings are finite without renormalization
and thus independent of the renormalization scheme (as long as it respects
BFM gauge invariance). Their asymptotic behavior is gauge-independent and
governed by the renormalization group. In particular, the
coefficients of the leading logarithms in the \ses\ are equal to the
ones appearing in the $\be$-functions associated with the running
couplings. All these properties are completely analogous to those of the
running coupling in QED; they follow in the same way from the relations
\refeq{eq:delZB} as in QED from
$Z_e = Z_{\Ahat\Ahat}^{-1/2}$. 

As mentioned above, the asymptotic
behavior of $e^2(q^2)$ and $g_2^2(q^2)$ is independent of the
quantum gauge parameter. The running couplings coincide in this
region with those defined in
\citeres{Ke89,Ku91,Si92}. For finite values of $q^2$, however,
there are differences%
\footnote{Those
differences also exist between the different
formulations of the previous treatments\cite{Ke89,Ku91,Si92}.},
and the couplings \refeq{eq:runcoupl} depend on $\xi_Q$.
This indicates that the mentioned desirable theoretical properties do
not single out any specific definition of the running couplings. Instead,
any definition of running couplings via
Dyson summation of \ses\ that take into account mass effects
is not unique but a matter of convention.
This arbitrariness is made transparent in the BFM
and has to be taken into account in treatments based on
running couplings.

\section{Non-linear Realization of the scalar Sector}
\label{se:nonlin}
In the previous sections the scalar $\mathrm{SU}(2)_{\rw}$ doublet $\hat\Phi$
was represented in the usual linear way, as defined in \refeq{eq:Hbq}. It is
interesting to inspect also the non-linear realization of the scalar
sector specified by \cite{nlhiggs,Je86}
\beq
\hat\Phi = \frac{1}{\sqrt2}(v+\hat H)\hat U,
\label{phinl}
\eeq
where the Goldstone fields $\hat\phi^a$ form the unitary matrix $\hat U$.
A convenient representation for $\hat U$ is for instance given by
\beq
\hat U = \exp\left(\frac{i}{v}\hat\phi^a\si^a\right).
\eeq
The $\hat\phi^a$ are related to the charge eigenstates $\hat\phi^\pm$,
$\hat\chi$
as follows
\beq
\hat\phi^\pm = \frac{1}{\sqrt{2}}\left(\hat\phi^2\pm i\hat\phi^1\right), \qquad
\hat\chi = -\hat\phi^3.
\eeq
The (physical) Higgs field $\hat H$ is a $\mathrm{SU}(2)_{\rw}$ singlet
unlike in
the linear parametrization
of \refeq{eq:Hbq}. The (non-polynomial) Higgs part of the
Lagrangian reads
\beq
{\cal L}_H =
\frac{1}{2}\tr{(\hat D_\mu\hat\Phi)^\dagger(\hat D^\mu\hat\Phi)}
+\frac{1}{2}\mu^2\tr{\hat\Phi^\dagger\hat\Phi}
-\frac{1}{16}\lambda\left(\tr{\hat\Phi^\dagger\hat\Phi}\right)^2,
\label{laghnl}
\eeq
where $\hat D_\mu$ denotes the covariant derivative of $\hat\Phi$ in matrix
notation
\beq
\hat D^\mu\hat\Phi =
\partial_\mu\hat\Phi-ig_2\hat W_\mu^a\frac{\si^a}{2}\hat\Phi
-ig_1\hat\Phi \hat B_\mu\frac{\si^3}{2}.
\eeq
One of the most interesting features of the non-linear realization
\refeq{phinl} is that the scalar self interaction in \refeq{laghnl} is
independent of the unphysical Goldstone fields $\hat\phi^a$ owing to the
unitarity of $\hat U$.
The linear and non-linear realizations of the scalar sector turn out to
be physically equivalent \cite{nlhiggs}, as the Jacobian of the corresponding
field transformation yields only a contribution to the Lagrangian proportional
to $\delta^{(D)}(0)$,
which cancels extra quartic UV divergences occurring in loop diagrams but
vanishes anyhow in dimensional regularization.

In the BFM the fields $\hat H$ and $\hat \phi^a$ are split into background and
quantum fields as follows \cite{HHint}
\beq
\hat H \to \Hhat + H, \qquad \hat U \to \Uhat U.
\eeq
Note that in order to preserve background gauge invariance the matrix $\hat U$
is split multiplicatively, i.e.\ the $\hat\phi^a$ are split in a non-linear
way.
The corresponding $R_\xi$-gauge-fixing term for the quantum fields reads
\cite{HHint}
\beqar
{\cal L}_{\rm GF} &=&
-\frac{1}{4\xi_Q} \tr{\left(
\partial^\mu W_\mu^a\si^a
+g_2\varepsilon^{abc}\What_\mu^a W^{\mu,b}\si^c
+\xi_Q\frac{g_2v}{2}\hat{U}\phi^a\si^a\hat{U}^\dagger\right)^2}
\nn\\[.3em] &&
{}-\frac{1}{2\xi_Q}\left(\partial^\mu B_\mu+\xi_Q\frac{g_1v}{2}\phi^3\right)^2,
\label{eq:gfterm}
\eeqar
and the Faddeev--Popov ghost Lagrangian ${\cal L}_{\rm FP}$ is
constructed as usual. Since ${\cal L}_{\rm GF}$ does not involve $H$ and
$\Hhat$, the physical Higgs field does not couple to the Faddeev--Popov
ghost fields.

Owing to the gauge invariance of the background Higgs field $\Hhat$,
vertex functions involving only $\Hhat$ fields are independent of
the gauge parameter $\xi_Q$. We have explicitly checked
this for the case of the tadpole $\Ga^{\Hhat}=iT^{\Hhat}$ and the Higgs
two-point function $\Ga^{\Hhat\Hhat}(q)=i(q^2-\MH^2)+i\Si^{\Hhat\Hhat}(q^2)$.
Hence, the tadpole counterterm $\delta t=-T^{\Hhat}$ and the Higgs-boson
mass counterterm $\delta\MH^2=\Re\left(\Si^{\Hhat\Hhat}(\MH^2)\right)$are
gauge-independent
in contrast to the corresponding quantities in the linear parametrization.
Moreover, the gauge independence of $\delta t$
implies the same for the gauge-boson mass counterterms $\delta\MW^2$ and
$\delta\MZ^2$
(and for the fermion-mass counterterms)
because of gauge independence of propagator poles.

Carrying out the field renormalization in a way respecting
background-field gauge invariance, one finds the
same relations, \refeq{eq:delZB}, for the field renormalization
constants as in the
linear scalar realization except for the one of $\delta Z_{\Hhat}$. There
is no constraint on $\delta Z_{\Hhat}$ following from gauge invariance.

In this context we mention that the non-polynomial scalar self
interactions in \refeq{laghnl} lead to a Higgs self-energy
$\Si^{\Hhat\Hhat}(q^2)$ which off-shell
 remains UV-divergent even after Higgs-field
and Higgs-mass renormalization. This is due to the presence of
UV-divergent terms proportional to $q^4$. Of course, in S-matrix
elements these spurious divergences always cancel against their
counterparts in other vertex functions
since the complete theory is renormalizable.

Disregarding the physical Higgs field in the non-linear realization
\refeq{phinl}, the SM reduces to the so-called gauged non-linear
$\si$-model (GNLSM) \cite{gnlsm}. The GNLSM is non-renormalizable but still
a $\mathrm{SU}(2)_{\rw}\times \mathrm{U}(1)_Y$ gauge theory.
The BFM effective action of the GNLSM
is gauge-invariant, and the corresponding vertex
functions obey simple Ward identities. However, the structure of these
Ward identities is different from the one in the SM described in the
previous sections, although they can be derived analogously. This is due
to the non-linearity in the scalar sector, which renders also gauge
transformations of the background Goldstone fields non-linear,
\beq
\delta\phihat^a = \MW\de\hat\theta^a
+\MW\frac{\sw}{\cw}\de\hat\theta^{\rm Y}\de^{a3}
-\frac{e}{2\sw}\varepsilon^{abc}\de\hat\theta^b\phihat^c
+\frac{e}{2\cw}\varepsilon^{a3c}\de\hat\theta^{\rm Y}\phihat^c
+ {\cal O}(\phihat^2),
\label{gaugatrnl}
\eeq
as can be easily
inferred from the detailed presentation of \citere{HHint}.
Consequently, a Ward identity for an $n$-point function in general
involves vertex functions with less external lines down to
self-energies.
Since $H$ and $\Hhat$ represent $\mathrm{SU}(2)_{\rw}\times \mathrm{U}(1)_Y$
singlets, the Ward identities of the GNLSM are valid in the SM with the
non-linear scalar realization
of \refeq{phinl}, too. The remaining Ward
identities in the SM
with non-linear scalar sector,
which involve $\Hhat$ vertex functions, are obtained
from the ones of the GNLSM simply by
taking further functional derivatives with
respect to $\Hhat$, or diagrammatically by adding further $\Hhat$ legs to
each occurring vertex function.
In particular, tadpole contributions can never occur in the Ward identities.
In \refeq{gaugatrnl}
the constant terms and the ones linear in the $\hat\phi^a$ coincide with
the corresponding result for the linear realization
of the scalar sector
[see Eq.\ (21) of \citere{bgflong}].
Therefore, Ward identities involving at most one Goldstone field but no
Higgs field in each occurring vertex function coincide within the linear
and non-linear scalar realizations.
In particular, this is the case for all Ward identities given in
\refse{sec:WI} except for \refeqs{eq:seZ2} and (\ref{eq:seW2}),
which are modified in the non-linear scalar realization of the SM and
the GNLSM to
\beqar
k^\mu \Ga^{\Zhat\chihat}_{\mu}(k) -i\MZ \Ga^{\chihat\chihat}(k)
&=& 0 , \\
k^\mu \Ga^{\What^\pm\phihat^\mp}_{\mu}(k)
\mp\MW \Ga^{\phihat^\pm\phihat^\mp}(k)
&=& 0,
\eeqar
where no tadpole contributions occur.

\section{Gauge Invariance and gauge-parameter-independent
Formulations of Green Functions}
\label{se:GIVvGIP}

In this section we
discuss the relation between gauge
invariance and gauge-pa\-ra\-me\-ter-independent formulations at the
level of Green functions.
One should be aware in this context that formally one can obtain a
gauge-parameter-independent quantity in a totally trivial way,
namely by putting the gauge parameters to a specific value,
e.g.~$\xi_i = 1$.
A ``trivial''
gauge-parameter independence of this kind obviously is not
related to any symmetry properties of the theory.

On the other hand, as mentioned in the
introduction, the rearrangement
of parts between different vertex functions in the conventional
formalism of the SM according to the prescription of the pinch technique
(PT) has led to new ``vertex functions'' that are
gauge-parameter-independent and coincide with the
corresponding vertex functions in the BFM for $\xiQ = 1$.
The PT ``vertex functions'' were found to fulfill the same
Ward identities which within the BFM are a direct consequence of gauge
invariance.

The origin of non-trivial symmetry relations in this case
stems from the fact that the gauge parameters in the
vertex functions are canceled while the lowest-order
propagators connecting the PT ``vertex functions'' are still
gauge-parameter-dependent. Obviously, this cannot be
achieved by simply putting the gauge parameters in the
conventionally defined
vertex functions
to a certain value.
As the complete S-matrix element is independent of the
gauge parameters, certain relations between the new ``vertex
functions'' must exist that enforce the cancellation of the
remaining gauge-parameter dependence.

It is important to note that the validity of non-trivial symmetry relations
is not based on the actual
gauge-parameter independence of the new ``vertex functions'',
but --- more generally --- on the independence of the gauge
parameters in the tree-level propagators from the gauge fixing
within loop diagrams. This, however, is exactly the same
situation as in the BFM. The vertex functions in the BFM depend
on the quantum gauge parameter $\xiQ$. This gauge dependence is
completely unrelated to the gauge fixing entering the
lowest-order propagators and giving rise to background gauge
parameters%
\footnote{In this section we restrict ourselves to linear background
gauge-fixing conditions. Note that the PT has only been formulated
for linear gauge fixings.}
$\xi_B^i$.
Thus, there is an analogy
between the BFM and prescriptions for constructing gauge-parameter-independent
``vertex functions'' in the conventional formalism,
as far as the cancellation of gauge-parameters associated with
lowest-order quantities is concerned.
In the BFM, however,
the cancellation of background gauge parameters is enforced by the BFM Ward
identities. Consequently,  a possible (and particularly simple)
choice for gauge-parameter-independent
``vertex functions'' constructed using the conventional formalism
is one that respects the BFM Ward identities.

In order to illustrate this in some more detail, we treat as a
simple example a four-fermion process
$\Pu_1\bar\Pd_1\to\Pu_2\bar\Pd_2$
at one-loop order,
where $\Pu_i$ and $\Pd_i$ are up- and down-type fermions, respectively.
For ease of notation we consider a
charged current process, i.e.\ we do not include mixing effects between
different gauge bosons.
The complete
one-loop contribution $\de {\cal M}$ to the
transition amplitude ${\cal M}$ can be written as
\beqar
\label{eq:4fermel}
\de {\cal M} &\rlap{=}& \phantom{{}+}
\left(\bar d_1 \Ga^{W^- \bar d_1 u_1}_{\mu,(0)} u_1\right)
\De^{W, \mu \al}
\left( \Ga^{W^+ W^-}_{\al\be,(1)}-
i\Ga^{W^+ W^-H}_{\al\be,(0)}\Ga^H_{(1)}/\MH^2\right)
\De^{W, \be \nu}
\left(\bar u_2 \Ga^{W^+ \bar u_2 d_2}_{\nu,(0)} d_2\right)
\no &&
{}+ \left(\bar d_1 \Ga^{W^- \bar d_1 u_1}_{\mu, (0)} u_1\right)
\De^{W, \mu \al}
\left( \Ga^{W^+ \phi^-}_{\al,(1)}
-i\Ga^{W^+ \phi^-H}_{\al,(0)}\Ga^H_{(1)}/\MH^2\right)
\De^{\phi}
\left(\bar u_2 \Ga^{\phi^+ \bar u_2 d_2}_{(0)} d_2\right)
\no &&
{}+ \left(\bar d_1 \Ga^{\phi^- \bar d_1 u_1}_{(0)} u_1\right)
\De^{\phi}
\left( \Ga^{\phi^+ W^-}_{\be,(1)}
-i\Ga^{\phi^+ W^-H}_{\be,(0)}\Ga^H_{(1)}/\MH^2\right)
\De^{W, \be \nu}
\left(\bar u_2 \Ga^{W^+ \bar u_2 d_2}_{\nu, (0)} d_2\right)
\no &&
{}+ \left(\bar d_1 \Ga^{\phi^- \bar d_1 u_1}_{(0)} u_1\right)
\De^{\phi}
\left( \Ga^{\phi^+\phi^-}_{(1)}
-i\Ga^{\phi^+\phi^-H}_{(0)}\Ga^H_{(1)}/\MH^2\right)
\De^{\phi}
\left(\bar u_2 \Ga^{\phi^+ \bar u_2 d_2}_{(0)} d_2\right)
\no &&
{}+ \left(\bar d_1 \Ga^{W^- \bar d_1 u_1}_{\mu, (1)} u_1\right)
\De^{W, \mu \nu} \left(\bar u_2 \Ga^{W^+ \bar u_2 d_2}_{\nu, (0)} d_2\right)
+ \left(\bar d_1 \Ga^{W^- \bar d_1 u_1}_{\mu, (0)} u_1\right)
\De^{W, \mu \nu} \left(\bar u_2 \Ga^{W^+ \bar u_2 d_2}_{\nu, (1)} d_2\right)
\no &&
{}+ \left(\bar d_1 \Ga^{\phi^- \bar d_1 u_1}_{(1)} u_1\right)
\De^{\phi} \left(\bar u_2 \Ga^{\phi^+ \bar u_2 d_2}_{(0)} d_2\right)
+ \left(\bar d_1 \Ga^{\phi^- \bar d_1 u_1}_{(0)} u_1\right)
\De^{\phi} \left(\bar u_2 \Ga^{\phi^+ \bar u_2 d_2}_{(1)} d_2\right)
\no &&
{}+ \bar d_1 \bar u_2 \Ga^{\bar d_1 u_1 \bar u_2 d_2}_{(1)} u_1 d_2,
\eeqar
where $\bar d_1, u_1, \bar u_2,$ and $d_2$ denote the spinors of the
external fermions. The subscripts ``(0)'' and ``(1)''
mark lowest-order and one-loop quantities, respectively.
The terms in the first four lines are \se\ and tadpole
contributions, the ones in the fifth and sixth line are vertex
corrections, and the last line contains the one-loop box contribution.
Since we are concerned with an S-matrix element,
\refeq{eq:4fermel} is understood to contain
{\it renormalized quantities} only.
In particular, the wave function renormalizations of the external
fermion lines are completely absorbed in the vertex corrections.

We use a linear $R_{\xi}$ gauge for the lowest-order propagators
$\De^{\phi}$ and $\De^{W}_{\mu \nu}$, i.e.
\beq
\label{eq:propzero}
\De^{\phi}(k) = \frac{i}{k^2 - \xi \MW^2} , \quad
\De^{W}_{\mu \nu}(k) = \left( - g_{\mu \nu} + \frac{k_{\mu}
k_{\nu}}{\MW^2} \right) \frac{i}{k^2 - \MW^2} - \frac{k_{\mu}
k_{\nu}}{\MW^2} \De^{\phi}(k) .
\eeq
According to our discussion above, we assume that
the gauge-parameter dependence of the one-particle
irreducible contributions in \refeq{eq:4fermel}
is not related to the
one of the tree propagators, \refeq{eq:propzero}.
This includes both the case of the BFM and of
gauge-parameter-independent ``vertex functions'' constructed in the
conventional formalism.

Since the box contribution is independent of the (background-type)
gauge parameter $\xi$, the cancellation of $\xi$
requires symmetry relations involving
self-energy and vertex contributions.
After inserting \refeq{eq:propzero} into \refeq{eq:4fermel}, the complete
$\xi$ dependence is contained in the term $\De^{\phi}$.
Collecting these gauge-dependent parts
yields two relations, namely one
for the contributions proportional to $\left( \De^{\phi}
\right)^2$ and one for the terms proportional to
$\De^{\phi}$.
Using the relations
\beq
\bar d_1 k^{\mu} \Ga^{W^- \bar d_1 u_1}_{\mu, (0)} u_1 =
\MW \bar d_1 \Ga^{\phi^- \bar d_1 u_1}_{(0)} u_1,
\qquad
\bar u_2 k^{\nu} \Ga^{W^+ \bar u_2 d_2}_{\nu, (0)} d_2 =
\MW \bar u_2 \Ga^{\phi^+ \bar u_2 d_2}_{(0)} d_2,
\eeq
and the tensor structure of the two-point functions, we find
\beqar
\label{eq:gaparrel1}
 k^{\al} k^{\be}  \Ga^{W^+ W^-}_{\al \be,(1)}(k)
 - 2 \MW k^{\al}  \Ga^{W^+ \phi^-}_{\al,(1)}(k)
 + \MW^2 \Ga^{\phi^+\phi^-,(1)}(k)
 -\frac{e\MW}{2\sw} \Ga^H_{(1)}
 &=& 0, \!
\hspace{2em} \\
\label{eq:gaparrel2}  \!
 \left( \bar d_1 k^{\mu} \Ga^{W^- \bar d_1 u_1}_{\mu, (0)} u_1 \right)
 2i\left[ \frac{k^{\al} k^{\be}}{k^2} \Ga^{W^+ W^-}_{\al \be,(1)}(k) -
 \frac{\MW k^{\al}}{k^2} \Ga^{W^+ \phi^-}_{\al,(1)}(k)
\right]
 \left(
   \bar u_2 k^{\nu} \Ga^{W^+ \bar u_2 d_2}_{\nu, (0)} d_2
 \right) && \no
{} +
 \left( \bar d_1 k^{\mu} \Ga^{W^- \bar d_1 u_1}_{\mu, (0)} u_1 \right)
 \frac{ie\MW}{\sw\MH^2}\Ga^H_{(1)}
 \left(
   \bar u_2 k^{\nu} \Ga^{W^+ \bar u_2 d_2}_{\nu, (0)} d_2
 \right) && \no
{} +
 \left(\bar d_1 k^{\mu} \Ga^{W^- \bar d_1 u_1}_{\mu, (0)} u_1 \right)
 \MW^2\left[
  \bar u_2 k^{\nu} \Ga^{W^+ \bar u_2 d_2}_{\nu, (1)} d_2
   - \MW \bar u_2 \Ga^{\phi^+ \bar u_2 d_2}_{(1)} d_2
 \right] && \no
{} \quad +
 \MW^2\left[
  \bar d_1 k^{\mu} \Ga^{W^- \bar d_1 u_1}_{\mu, (1)} u_1
   - \MW \bar d_1 \Ga^{\phi^- \bar d_1 u_1}_{(1)} u_1
 \right]
 \left(\bar u_2 k^{\nu} \Ga^{W^+ \bar u_2 d_2}_{\nu, (0)} d_2 \right)
 &=& 0, \!
\eeqar
where $k^\mu$ represents the
total incoming momentum of the initial state.
\refeq{eq:gaparrel1}
coincides with
the (renormalized) Ward
identity valid for the one-loop \ses\ in the conventional formalism of
the SM, while \refeq{eq:gaparrel2} involves both process-specific
vertex contributions and process-independent \ses\ and a tadpole term.
Note that the \se\ and vertex contributions in
\refeq{eq:gaparrel2} do not necessarily decouple.

In the particular case of the BFM with the renormalization procedure
described in \refse{se:ren}, \refeqs{eq:gaparrel1} and
\refeqf{eq:gaparrel2} are obviously fulfilled.
\refeq{eq:gaparrel1} is just the sum of the BFM Ward
identities \refeqs{eq:seW1} and \refeqf{eq:seW2}.
In \refeq{eq:gaparrel2} the four lines
actually vanish separately.
The first line is zero owing to the Ward identity \refeq{eq:seW1}, the
second is absent since the tadpole is renormalized to zero, and the last
two lines vanish owing to the Ward identities \refeq{WIWff}
and the on-shell conditions for the fermions.

In this context, it is interesting to add some remarks on the tadpole
contributions. Of course, it is not necessary to renormalize the tadpole
to zero as it is done in \refse{se:ren}. Instead, one can fix its
renormalized value arbitrarily or one need not renormalize it at all.
This leads to additional
tadpole contributions in all mass counterterms, e.g.
\beqar
\label{eq:dMW2}
\de\MW^2 &=& \Re\left(\Si^{WW}_{0,\rm T}(\MW^2)\right)
-\frac{e\MW}{\sw\MH^2}T^H, \\
\de\Mf   &=& \frac{1}{2}\Mf \Re\Bigr[\Si^{\bar ff}_{0,\rL}(\Mf^2)
                                     + \Si^{\bar ff}_{0,\rR}(\Mf^2)
                                     + 2\Si^{\bar ff}_{0,\rS}(\Mf^2)
                 \Bigl]
-\frac{e\,\Mf}{2\sw\MW\MH^2}T^H,
\label{eq:dmtad}
\eeqar
where the unrenormalized self-energies $\Si_0$
are defined like in \citere{bgflong}.
The renormalized tadpole $T^H=T^H_0+\de t$
consists of the unrenormalized tadpole
contribution $T^H_0$ and the tadpole counterterm $\de t$.
The tadpole terms in \refeqs{eq:dMW2} and \refeqf{eq:dmtad} are canceled in
$\de {\cal M}$ by the tadpole contributions in \refeq{eq:4fermel}.
Consequently, in such a renormalization scheme the four lines in
\refeq{eq:gaparrel2} do {\it not} decouple. Using the BFM with a finite
renormalized tadpole, the situation is as follows:
the first line of
\refeq{eq:gaparrel2} is still zero owing to the Ward identity
\refeq{eq:seW1}. However, the last two lines yield finite tadpole
contributions upon inserting the identities \refeqs{WIWff}
and using the on-shell conditions for the fermions,
\beq
{\mathrm Re}\left.\left\{ \Ga^{\Ffbar f}_{(1)}(-p)
- \frac{i}{\MH^2} \Ga^{\Ffbar f \FH}_{(0)}(-p,p,0)
\Ga^{\FH}_{(1)} \right\} \right|_{p^2 = \Mf^2} u(p) = 0.
\eeq
The resulting
terms cancel exactly against the tadpole contribution in \refeq{eq:gaparrel2}.

The above investigation of the gauge-parameter dependence associated
with the tree lines for the example of a (charged-current) four-fermion
process shows, in particular, that the gauge independence of the
corresponding S-matrix element does not require a decoupling of the
conventional Ward identity \refeq{eq:gaparrel1} into the BFM
counterparts \refeqs{eq:seW1} and \refeqf{eq:seW2}. This is in contrast to
the statements made in \citere{PT3} in the PT framework.
There the
decoupled Ward identities were derived under the {\it additional assumption}
that the tree-like gauge-parameter dependence of self-energy
contributions is canceled independently of the remaining
vertex and tadpole contributions.
In particular, care has to be taken with respect to the tadpole
contributions. They cannot be simply included in the self-energies
obeying the decoupled identities \refeqs{eq:seW1} and \refeqf{eq:seW2},
since they do not fulfill these identities
by
themselves. Finally, we emphasize that derivations starting from the gauge
independence of the S-matrix can only yield results for {\it renormalized}
vertex functions since an ``unrenormalized S-matrix''
does not exist.

Even if gauge-parameter-independent ``vertex functions'' are constructed
in such a way that they fulfill the BFM Ward identities, their
definition is still not unique.
One can always shift parts between the ``vertex functions'' that by
themselves fulfill the Ward identities.
This freedom naturally appears within the framework of the BFM as the
freedom of choosing different values of the quantum gauge parameter
$\xiQ$. As has been stressed above, the BFM Ward identities and the
desirable properties of the BFM vertex functions are a
consequence of gauge invariance and hold for arbitrary
values of $\xiQ$.

When comparing the approach pursued e.g.~in the PT with the BFM,
one should keep in mind that the field-theoretical
interpretation of the PT quantities being defined by a
rearrangement of contributions between different vertex
functions is rather obscure. Their process independence has not
been proven in general,
but only verified for specific examples (see in particular \citere{PT2}),
and their construction beyond one-loop
order is technically very complicated~\cite{twoloop}. In
contrast, the BFM vertex functions have a well-defined
field-theoretical meaning and can be derived from the effective
action in all orders of perturbation theory.

As we have seen above, in the conventional formalism the application of
the PT is a special case of decoupling the gauge-parameter
dependence of the vertex functions from the one in the tree
propagators.
Recently,
however, the PT has also been applied within the framework of
the BFM in order to eliminate the dependence of the BFM vertex functions
on the quantum gauge parameter $\xiQ$~\cite{papavBFM}.
Since the background gauge parameters $\xi_B$ appearing in the
tree propagators of the BFM are not related to $\xiQ$, an
elimination of $\xiQ$ via a prescription like the
PT can {\em not} be distinguished from trivially putting $\xiQ$ to any
specific value. This can also be seen from the fact that
application of the PT within the BFM does not lead to new relations
between the BFM vertex functions. The apparent gauge-parameter
independence has only been achieved on cost of the specific prescription
used in the PT to eliminate the gauge parameter.

The comparison between PT and BFM has made transparent
that, despite their gauge-parameter independence and several
desirable properties, the PT ``vertex functions'' are not unique
but to a large extent a matter of convention.
This is evident, because any off-shell quantity cannot be
directly related to an observable and thus cannot uniquely be fixed.
It was already pointed out in \citere{Je86} that off-shell quantities
are ambiguous even if gauge invariance is imposed. This holds in
particular for all off-shell form factors such as a neutrino
electromagnetic moment or  anomalous triple-gauge-boson couplings.
While these quantities are well-defined where the single one-particle
exchange approximation holds, like e.g.\ on the $\PZ$ resonance, they are
not directly observable in general and to a large extent ambiguous.
The PT, like any other prescription, can only provide a more or less
convenient definition for off-shell quantities but cannot supply a
physical meaning.

\section{The $S$, $T$, and $U$ Parameters in the
Background-Field Method}
\label{se:STU}

As an illustration of the discussion given in the last section,
we treat the $S$, $T$, and $U$ parameters in the framework
of the BFM.
The $S$, $T$, and $U$ parameters are defined as certain
combinations of \ses~\cite{STU}. Originally,
they were introduced in order
to parametrize the effects of new physics that
enters only via oblique (i.e.~\se)
corrections. They can be extracted from
experiment by comparing 
the experimentally measured values ${\cal A}_i^{\mathrm{exp}}$
of a number of observables
with their values predicted by the SM, ${\cal A}_i^{\mathrm{SM}}$,
i.e.
\beq
\label{eq:STUgen}
{\cal A}_i^{\mathrm{exp}} = {\cal A}_i^{\mathrm{SM}} +
f^{\mathrm{NP}}_i(S, T, U).
\eeq
Here ${\cal A}_i^{\mathrm{SM}}$ contains the complete
radiative corrections in the SM up to a given order, while
$f^{\mathrm{NP}}_i(S, T, U)$ is a function of the parameters
$S$, $T$, $U$ and describes the contributions of new
physics. The SM prediction ${\cal A}_i^{\mathrm{SM}}$ is
evaluated for a reference value of $m_\Pt$ and $M_\PH$. For most
observables accessible by precision measurements
the corrections caused by a variation of $m_\Pt$ and
$M_\PH$ can also be absorbed into the parameters $S$, $T$,
and $U$.

The parameters $S$, $T$, and $U$ obtained via \refeq{eq:STUgen}
are gauge-invariant quantities. This follows from the
fact that ${\cal A}_i^{\mathrm{SM}}$ contains a complete set
of electroweak radiative corrections entering an S-matrix
element and that
the analysis has been restricted to those models of
new physics where $f^{\mathrm{NP}}_i(S, T, U)$ accounts for the
total contribution.

In \citere{DKS}, however, an extension 
of the $S$, $T$, and $U$ parametrization to cases where these
assumptions do not hold
has been discussed.
This includes effects of new physics that do not
exclusively enter via oblique corrections but also via vertex
and box contributions as for example
anomalous triple-gauge-boson couplings. Furthermore, the authors
of \citere{DKS} also considered the case where the $S$, $T$, and
$U$ parameters are used {\em within} the SM,
i.e.~to parametrize not only new physics effects but also
the SM fermionic and bosonic radiative corrections.

These extensions of the $S$, $T$, and $U$ parameters appear to be
questionable, since the parameters defined in this way are no longer
directly related to
observables. In particular, this poses severe
problems of gauge invariance.
It was pointed out in \citere{DKS} that
calculating the one-loop bosonic SM
contributions to the $S$, $T$, and $U$ parameters
yields gauge-parameter-dependent results. It was further noted
that for gauges
with $\xi_W \neq \cw^2 \xi_Z + \sw^2 \xi_A$ the parameters
$T$ and $U$ are even UV-divergent.
The authors of \citere{DKS} argued
that these problems can be overcome
by using the PT in order to eliminate the gauge-parameter
dependence of the one-loop gauge-boson \ses. By explicit
calculation, the $S$, $T$, and $U$ parameters obtained within
the PT were also shown to be UV-finite.

In order to discuss the formulation of the $S$, $T$, and $U$
parameters given in \citere{DKS}, we calculate the bosonic SM contributions
to the $S$, $T$, and $U$ parameters in the framework of the BFM.
To allow for an easy comparison,
we adopt the same definition of $S$, $T$, and $U$ as in \citere{DKS}, i.e.
\beqar
\al S_0 &=&
4\cw^2\sw^2 {\mathrm Re}\left\{
-\Pi^{ZZ}_0(\MZ^2) + \frac{\sw^2-\cw^2}{\cw\sw}\Pi^{\ga Z}_0(\MZ^2)
+\Pi^{\ga\ga}_0(\MZ^2) \right\}, \\
\al T_0 &=&
-\frac{\Si^{WW}_{\rm T,0}(0)}{\MW^2} + \frac{\Si^{ZZ}_{\rm T,0}(0)}{\MZ^2}
-2\cw\sw\frac{\Si^{\ga Z}_{\rm T,0}(0)}{\MW^2}, \\
\al U_0 &=&
4\sw^2 {\mathrm Re}\left\{
-\Pi^{WW}_0(\MW^2) + \cw^2\Pi^{ZZ}_0(\MZ^2)
-2\cw\sw\Pi^{\ga Z}_0(\MZ^2) + \sw^2\Pi^{\ga\ga}_0(\MZ^2) \right\},
\hspace{2em}
\eeqar
where as usual $\al = e^2/(4 \pi)$. We use the subscript ``0'' to
indicate that $S$, $T$, and $U$ are defined in terms of
unrenormalized one-loop \ses.
Note that in our conventions $\sw$ differs by a sign from the one 
used in \citere{DKS}. Furthermore, we use the on-shell definitions for
$e$ and $\sw$, while in \citere{DKS} the $\overline{\mbox{MS}}$ parameters
are used. This difference is irrelevant for the discussion in this
section.

The bosonic contributions to $S$, $T$ and $U$ in the BFM formulation of the
SM read
\beqar
\lefteqn{\! \! \!
\al S^{\mathrm{SM, BFM}}_0 = \frac{\al}{24 \pi} \biggl\{
 2 \cw^2 - 5 + h + 2 \cw^2 \xiQ
 - 2 \log(\cw^2) - 2 \left[3 - (1 + 2 \cw^2) \xiQ \right]
  \frac{\log(\xiQ)}{1 - \xiQ} } \no
&& {} - 2 (12 - 4 h + h^2) F(\MZ^2; \MH, \MZ)
+ 6 (1 - 4 \cw^2 \xiQ) F(\MZ^2; \sqrt{\xiQ} \MW, \sqrt{\xiQ} \MW)
\no
&& {} - 4 \left[1 + 10 \cw^2 + \cw^4
  - 2 \cw^2 (1 + \cw^2) \xiQ
  + \cw^4 \xiQ^2 \right] F(\MZ^2; \MW, \sqrt{\xiQ} \MW)
\biggr\},
\label{eq:SBFM} \\[0.2 cm]
\lefteqn{\! \! \!
\al T^{\mathrm{SM, BFM}}_0 = \frac{\al}{16 \pi} \biggl\{
 \frac{1}{\sw^2} (12 - 5 \xiQ)
- \frac{3 h^2}{\cw^2 (1 - h) (\cw^2 - h)}
  \log (h) } \no
&& {}
 + \frac{\cw^2\log(\cw^2)}{\sw^4 (\cw^2 - h) (\cw^2 - \xiQ)
 (1 - \cw^2 \xiQ)}
\Bigl[3 \cw^2 (3 - 2 \cw^2 + 3 \cw^4)
 - 3 h (2 - \cw^2 + 3 \cw^4)  \no
&& {} \; \;  - \Bigl(9 - 12 \cw^2 + 23 \cw^4 + 9 \cw^6
 - h (6/\cw^2 - 9 + 20 \cw^2 + 12 \cw^4) \Bigr) \xiQ \no
&& {} \; \;  + \Bigl(\cw^2 (5 + 6 \cw^2 + 11 \cw^4)
           - h (2 + 9 \cw^2 + 11 \cw^4) \Bigr) \xiQ^2
- \cw^2 (2 + 3 \cw^2) (\cw^2 - h) \xiQ^3 \Bigr]
\no
&& {} + \frac{3 \left[
  3 \cw^2 - (3 + 2 \cw^2) \xiQ + (1 + 2 \cw^2) \xiQ^3 - \cw^2 \xiQ^4
 \right]}{(\cw^2 - \xiQ) (1 - \cw^2 \xiQ)}
 \frac{\log(\xiQ)}{1 - \xiQ} \biggr\} , \label{eq:TBFM}\\[0.2 cm]
\lefteqn{\! \! \!
\al U^{\mathrm{SM, BFM}}_0 = \frac{\al}{12 \pi \cw^2} \biggl\{
 - \frac{2}{\cw^2} - \frac{39}{2} + \frac{171}{2} \cw^2 - \cw^4
 + \frac{1}{2} h \sw^2 + (4 + 12 \cw^2 - \cw^4) \xiQ } \no
&& {} + \frac{\cw^4\log(\cw^2)}{\sw^2 (\cw^2 - \xiQ) (1 - \cw^2 \xiQ)}
 \Bigl[ 1 + 89 \cw^2 - 27 \cw^4
- (1/\cw^2 + 107 - 41 \cw^2 + 44 \cw^4) \xiQ
\no && {} \; \;
 + (13 + 53 \cw^2 - 33 \cw^4) \xiQ^2
+ 3 \cw^2 (2 + 3 \cw^2) \xiQ^3 \Bigr] \no
&& {} - \frac{\sw^2}{(\cw^2 - \xiQ) (1 - \cw^2 \xiQ)}
 \Bigl[ 1 + 5 \cw^2 + 27 \cw^4
 - (1/\cw^2 + 4 + 15 \cw^2 + 25 \cw^4) \xiQ \no
&& {} \; \;  - (1/\cw^2 + 12 - 6 \cw^2 + 13 \cw^4 - 2 \cw^6) \xiQ^2
\no && {} \; \;
+ (1 + 22 \cw^2 + 16 \cw^4) \xiQ^3
 - 9 \cw^4 \xiQ^4 \Bigr] \frac{\log(\xiQ)}{1 - \xiQ} \no
&& {} - (12 \cw^2 - 4 h + h^2/\cw^2) F(\MW^2; \MH, \MW)
 + \cw^2 (12 - 4 h + h^2) F(\MZ^2; \MH, \MZ)
\no && {} + 48 \cw^2 \sw^2 F(\MW^2; 0, \MW)
+ \frac{\sw^2}{\cw^2} (1 + 5 \cw^2) (1 - 4 \cw^2 \xiQ)
 F(\MZ^2; \sqrt{\xiQ} \MW, \sqrt{\xiQ} \MW)
\no && {}
- 2 \sw^2 (1 + \cw^2) \Bigl[1/\cw^2 + 10 + \cw^2 - 2 (1 + \cw^2) \xiQ
 + \cw^2 \xiQ^2 \Bigr] F(\MZ^2; \MW, \sqrt{\xiQ} \MW)
\no && {}
-  (1 - 4 \cw^2) (1/\cw^2 + 20 + 12 \cw^2)
 \left[F(\MW^2; \MW, \MZ) - F(\MZ^2; \MW, \MW) \right]
\biggr\} , \label{eq:UBFM}
\eeqar
where we have used the shorthand $h = \MH^2/\MZ^2$,
and the quantum gauge parameter $\xiQ = \xiQ^W = \xiQ^B$ has been kept
as a free parameter. The UV-finite
function $F(p^2; m_1, m_2)$ is defined as
\beq
F(p^2; m_1, m_2) = {\mathrm Re} \left(
  B_0(p^2; m_1, m_2) - B_0(0; m_1, m_2) \right) ,
\eeq
where $B_0(p^2;m_1,m_2)$ is the usual scalar one-loop two-point
integral~\cite{Dehab}.
For completeness, we also give the difference between $S^{\mathrm{SM,
BFM}}_0$, $T^{\mathrm{SM, BFM}}_0$, $U^{\mathrm{SM, BFM}}_0$ evaluated at
$\xiQ = 1$ and the bosonic contributions to the $S$, $T$, and $U$
parameters calculated in the 't~Hooft--Feynman gauge (tHF) of the
conventional formalism:
\beqar
\al S^{\mathrm{SM, BFM}}_0\biggr|_{\xiQ = 1}  &=&
  \al S^{\mathrm{SM, conv}}_0\biggr|_{\mathrm{tHF}} +
\frac{2\al\cw^2}{\pi}
{\mathrm Re}\left\{B_0(0,\MW,\MW)-B_0(\MZ^2,\MW,\MW)\right\},
\nn\\
\al T^{\mathrm{SM, BFM}}_0\biggr|_{\xiQ = 1}  &=&
  \al T^{\mathrm{SM, conv}}_0\biggr|_{\mathrm{tHF}} +
\frac{\al}{\sw^2\pi}\left\{
B_0(0,\MW,\MW)-\sw^2 B_0(0,0,\MW)
\right. \nn\\ && \hspace{10em}\left.
{}-\cw^2 B_0(0,\MW,\MZ) \right\}, \nn\\
\al U^{\mathrm{SM, BFM}}_0\biggr|_{\xiQ = 1}  &=&
  \al U^{\mathrm{SM, conv}}_0\biggr|_{\mathrm{tHF}} +
\frac{4\al}{\pi}{\mathrm Re}\left\{
\sw^2 B_0(0,0,\MW)-\cw^2 B_0(0,\MW,\MW)
\right. \nn\\ && \hspace{6em}\left.
{}+\cw^2 B_0(0,\MW,\MZ) -\sw^2 B_0(\MZ^2,\MW,\MW)\right\}.
\hspace{2em}
\eeqar
This coincides with the result obtained within the PT given in \citere{DKS}.

As can be seen in \refeqs{eq:SBFM}, (\ref{eq:TBFM}) and (\ref{eq:UBFM}),
$S^{\mathrm{SM, BFM}}_0$, $T^{\mathrm{SM, BFM}}_0$, and
$U^{\mathrm{SM, BFM}}_0$ are UV-finite for
arbitrary values of $\xiQ$. While within the PT the UV-finiteness of
the parameters could only be inferred from explicit computation, in
the BFM it is an obvious consequence of gauge invariance%
\footnote{Note that in the BFM gauge invariance restricts
the number of quantum gauge parameters to two, $\xiQ^W$ and
$\xiQ^B$. This automatically implies the identity $\xiQ^W =
\cw^2 \xiQ^Z + \sw^2 \xiQ^A$.}.
In order to show this, we consider the renormalized 
$S$, $T$ and $U$ parameters. The renormalization of $T^{\mathrm{SM,
BFM}}_0$, for instance, yields
\beqar
\al T^{\mathrm{SM, BFM}} &=&
 - \frac{\Si_{\rT}^{\What\What}(0)}{\MW^2}
 + \frac{\Si_{\rT}^{\Zhat\Zhat}(0)}{\MZ^2}
\no &=&
- \frac{\Si_{\rT,0}^{\What\What}(0)}{\MW^2} + \delta Z_{\What} +
  \frac{\delta \MW^2}{\MW^2} + \frac{\Si_{\rT,0}^{\Zhat\Zhat}(0)}{\MZ^2}
  - \delta Z_{\Zhat\Zhat} - \frac{\delta \MZ^2}{\MZ^2} ,
\eeqar
where we have used that in the BFM
$\Si_{\rT,0}^{\Ahat\Zhat}(0) = \Si_{\rT}^{\Ahat\Zhat}(0) = 0$ holds,
which can be inferred from \refeq{eq:sega}.
However, from \refeq{eq:delZB} we find
\beq
\delta Z_{\What} + \frac{\delta \MW^2}{\MW^2} - \delta Z_{\Zhat\Zhat} -
\frac{\delta \MZ^2}{\MZ^2} = 0 ,
\eeq
and therefore
\beq
\label{eq:Tren}
\al T^{\mathrm{SM, BFM}} = \al T^{\mathrm{SM, BFM}}_0 =
- \frac{\Si_{\rT}^{\What\What}(0)}{\MW^2} +
\frac{\Si_{\rT}^{\Zhat\Zhat}(0)}{\MZ^2} .
\eeq
Since $T^{\mathrm{SM, BFM}}_0$ and $T^{\mathrm{SM, BFM}}$ are identical,
the unrenormalized parameter $T^{\mathrm{SM, BFM}}_0$ is manifestly
UV-finite.
Similarly one derives
\beq
\al S^{\mathrm{SM, BFM}} = \al S^{\mathrm{SM, BFM}}_0 , \quad
\al U^{\mathrm{SM, BFM}} = \al U^{\mathrm{SM, BFM}}_0 .
\eeq

For fermionic contributions, the combination of \ses\ appearing in
\refeq{eq:Tren} is just the one-loop correction to the $\rho$ parameter.
While the bosonic contributions to this combination of \ses\ are
divergent in the
conventional formalism of the SM, they are finite within the BFM.

Recalling the discussion of the previous section, it should now be
obvious that the definition of the $S$, $T$, and $U$ parameters
based on the PT is not distinguished, neither through
its UV-finiteness nor through its apparent gauge-parameter
independence.
This ambiguity 
reflects the fact that a parametrization of the SM
bosonic contributions in terms of $S$, $T$, and $U$ cannot directly be
compared to experimentally measured quantities. Moreover, there is
a priori no reason why the $S$, $T$, and $U$ parameters defined within
the PT should include the dominant part of the bosonic contributions
to electroweak observables.
In fact, comparing for the bosonic contributions
the complete one-loop result of the $\rho$ parameter stated
in \citere{Degrass} with the PT value of $\al T$~\cite{DKS},
one finds that the process-specific bosonic contributions that
are not included in the PT definition of $\al T$ give by far the
most important contribution. The bosonic PT result
even has a sign different from
the complete bosonic one-loop contribution
to the $\rho$ parameter.%
\footnote{We have assumed an electron target and varied the
Higgs mass between 50 and 1000 GeV.}
Furthermore, from the analysis
of LEP1 observables and muon decay carried out in \citere{Schild2}
it can directly be seen that the (universal) bosonic corrections associated
with the PT gauge-boson self-energies in general do {\it not} represent
the dominant bosonic effects.

In summary, while well established for the treatment of new physics
contributions entering solely via vacuum polarization effects,
the framework of the $S$, $T$, and $U$ parameters appears not to be
favorable for an incorporation of SM bosonic corrections or of new
physics effects going beyond oblique corrections.
As we have seen, their definition becomes ambiguous in these cases.
In order to study the complete SM contributions,
it seems to be more appropriate to directly
inspect observables or (process-specific) effective parameters
closely related to measurable quantities. For LEP1 physics such
parametrizations were e.g.~proposed in \citere{Alt} and
\citeres{Schild1,Schild2}.

\section{Conclusion}
\label{concl}
Quantizing a gauge theory within the background-field method (BFM)
yields a
manifestly gauge-invariant effective action for the underlying model.
The application of this method
to the electroweak Standard Model has been reviewed and
further investigated. We have derived consequences of the
simple Ward identities that follow directly from gauge invariance
of the effective action. In particular, we have discussed the impact of
BFM gauge invariance on renormalization.
Moreover, we have considered the generalization of the
BFM to the non-linear realization of the scalar
sector of the Standard Model.

The interplay between  gauge-parameter independence of the S-matrix
and Ward identities relating vertex functions has been further explored.
We have shown that any formalism that decouples the
gauge-parameter dependence of the vertex functions from the one of the
tree lines leads to symmetry constraints
for the corresponding ``vertex functions''.
These quantities are, however, not uniquely determined by this requirement,
but it is possible to shift parts between ``vertex functions''
that by themselves obey the constraints.
This fact signals the ambiguity
which within the BFM is naturally made transparent by
the dependence of the vertex functions on the quantum gauge parameter.

Although approaches based on a redistribution of parts between different
Green functions may yield ``vertex functions'' that coincide with the
corresponding quantities in the BFM, from a conceptual
point of view these methods differ considerably.
In addition to being technically rather complicated, approaches
like the pinch technique suffer from severe theoretical shortcomings.
In particular, the
field-theoretical meaning of objects constructed by redistributions is not
clear. In contrast, the BFM vertex functions have a well-defined
field-theoretical interpretation and are derived from an effective
action in all orders of perturbation theory.

The application of a gauge-parameter elimination procedure within the
BFM degenerates to a trivial selection of a particular value for the
quantum gauge parameter and thus to a mere convention.

Since off-shell quantities such as Green functions are not directly related
to observables, they cannot be fixed on physical grounds.
Therefore, any
prescription that fixes these quantities can only be a more
or less convenient definition but cannot be unique. We have illustrated
this fact by calculating and discussing the (gauge-dependent) standard
contributions to the $S$, $T$, and $U$ parameters.

\vspace{1em}
\pagebreak
\noindent
{\bf Note added}

We would like to comment on remarks made in
\citere{papanew} concerning the connection between background-field
method and pinch technique. There, \citere{BFMvPT} was cited in the
context of the ``erroneous impression'' and the ``naive expectation''
that ``Green's functions calculated within the background-field method
should be completely gauge-invariant, and identical to the
corresponding pinch-technique Green's functions''.
Furthermore, with respect to the gauge-pa\-ra\-me\-ter dependence of the
background-field vertex functions, it was stated in \citere{papanew}
that ``there'' (\citere{BFMvPT}) ``was an attempt to assign a physical
significance to this dependence''. {\it None} of these statements
has been made in \citere{BFMvPT}, where all statements and
conclusions are based on facts but not on the (irrelevant) ``initial
expectations'' mentioned in \citere{papanew}. Note that
one of our conclusions was that the gauge-parameter dependence in the
BFM signals the fact that it is {\em not} possible to assign a physical
significance to off-shell Green functions.

\vspace{-0.2em}
\section{Acknowledgements}
\vspace{-.5em}
S.D.\ would like to thank Carsten Grosse-Knetter for many helpful discussions
on the non-linear representations of scalar fields and
for a fruitful collaboration.

\end{document}